\documentclass[preprint,showpacs,showkeys,preprintnumbers,amsmath,amssymb,superscriptaddress]{revtex4}
\usepackage{mathrsfs}
\usepackage[sort&compress]{natbib}
\usepackage{graphicx}
\usepackage{dcolumn}
\usepackage{bm}
\usepackage{subfigure}
\usepackage{picinpar}
\usepackage{picins}
\usepackage{epic}
\usepackage{psfrag}
\begin{document}
\preprint{}
\title{Complex dynamics of a Holling-type IV predator-prey model}
\author{Lei Zhang}
\affiliation{Institute of Nonlinear Analysis, School of Mathematics
and Information Science, Wenzhou University, Wenzhou,Zhejiang,
325035} \affiliation{Department of Mathematics, North University of
China, Taiyuan, Shan'xi 030051, P.R. China}
\author{Weiming Wang}
\email{weimingwang2003@163.com}  \affiliation{Institute of Nonlinear
Analysis, School of Mathematics and Information Science, Wenzhou
University, Wenzhou,Zhejiang, 325035}
\author{Yakui Xue}
\affiliation{Department of Mathematics, North University of China,
Taiyuan, Shan'xi 030051, P.R. China}
\author{Zhen Jin}
\affiliation{Department of Mathematics, North University of China,
Taiyuan, Shan'xi 030051, P.R. China}

\date{\today}

\begin{abstract}
In this paper, we focus on a spatial Holling-type IV predator-prey
model which contains some important factors, such as diffusion,
noise (random fluctuations) and external periodic forcing. By a
brief stability and bifurcation analysis, we arrive at the Hopf and
Turing bifurcation surface and derive the symbolic conditions for
Hopf and Turing bifurcation in the spatial domain. Based on the
stability and bifurcation analysis, we obtain spiral pattern
formation via numerical simulation. Additionally, we study the model
with colored noise and external periodic forcing. From the numerical
results, we know that noise or external periodic forcing can induce
instability and enhance the oscillation of the species, and resonant
response. Our results show that modeling by reaction-diffusion
equations is an appropriate tool for investigating fundamental
mechanisms of complex spatiotemporal dynamics.
\end{abstract}

 \maketitle

\section{Introduction}

Predation, a complex natural phenomena, exists widely in the world,
e.g., the sea, the plain, the forest, the desert and so
on~\cite{Cantrell2003}. To model this phenomenon, predator-prey
model has been suggested for a long time since the pioneer work of
Lotka and Volterra~\cite{Kendall}. The predator-prey model is a kind
of ``pursuit and evasion" system in which the prey try to evade the
predator and the predator try to catch the prey if they
interact~\cite{Murray2003}. Pursuit means the predator try to
shorten the spatial distance between the predator and the prey,
while evasion means the prey try to widen this spatial distance. In
fact, the predator-prey model is a mathematical method to
approximate some part of our real world. And the dynamic behavior of
the predator-prey model has long been and will continue to be one of
the dominant themes in both ecology and mathematical ecology due to
its universal existence and
importance~\cite{Berryman,kuang98global}.

Generally, a classical predator-prey model can be written as the
form~\cite{Arditi,David2002}:
\begin{eqnarray}\label{eq:1}
\begin{array}{l}
 \frac{dN}{dt}=Nf(N)-mPg(N,P),\qquad
 \frac{dP}{dt}=P[cmg(N,P)-d],
\end{array}
\end{eqnarray}
where $N$ and $P$ stand for prey and predator quantity,
respectively, $f(N)$ the per capita rate of increase of the prey in
absence of predation, $d$ the food-independent death rate of
predator, $g(N,P)$ the functional response, the prey consumption
rate by an average single predator, which obviously increases with
the prey consumption rate and can be influenced by the predator
density, and which refers to the change in the density of prey
attached per unit time per predator as the prey density changes,
$mg(N, P)$ the amount of prey consumed per predator per unit time,
and $cmg(N,P)$ the predator production per capita with predation.

In population dynamics, a functional response $g(N, P)$ describes
the relationship between the predator and their prey, and the
predator-prey model is always named after the corresponding
functional response for its key
position~\cite{Ruan,Abrams,Arditi,David2002}. In the history of
population ecology, both ecologists and mathematicians have a great
interest in the Holling-type predator-prey
models~\cite{Pang,Jicai2004,Ko,Weipeng,
Gakkhar,Yuming2004,Abrams,Estep,Skalski,Teemu2004,Malchow2004,
Murray2003,HUAIPING2002,ZhangChen}, including Holling type I-III,
originally due to Holling~\cite{Hollinga,Hollingb}, and Holling type
IV, suggested by Andrews~\cite{Andrews}. The Holling-type functional
responses are so-called ``prey-dependent" type~\cite{Abrams}, for
$g(N,P)$ in Eq.\eqref{eq:1} is a function only related to prey $N$.
The classical expression of Holling-type II functional response is
$g(N, P)=\frac {mN}{1+b{N}}$, and $g(N, P)=\frac {mN^2}{1+a{N}^{2}}$
is called Holling-type III. The Holling-type IV functional response
is written as follows:
\begin{equation}\label{eq05}
g(N, P)=\frac {mN}{1+bN+a{N}^{2}}.
\end{equation}
Fuction~\eqref{eq05} is called Monod-Haldane-type functional
response too~\cite{Karen}. In addition, when $b=0$, a simplified
form $ g(N, P)=\frac {mN}{1+a{N}^{2}}$ is proposed by Sokol and
Howell~\cite{Sokolhowell}, and some scholars also called it as
Holling-type-IV~\cite{Karen,Ruan}. In this paper, we focus on the
Holling-type IV functional response which takes the
form~\eqref{eq05}, and the corresponding predator-prey model which
takes the form:
\begin{eqnarray}\label{eq07}
 \begin{array}{l}
 \frac{d N}{d t}=rN \left( 1-{\frac {N}{K}} \right)
  -{\frac {mNP}{1+bN+a{N}^{2}}},\qquad
 \frac{d P}{d t}=P \left( -q+{\frac {cmN}{1+bN+a{N}
 ^{2}}} \right),
 \end{array}
\end{eqnarray}
where $r>0$ stands for maximum per capita growth rate of the prey,
$m>0$ the capture rate, $c>0$ the conversion rate of prey captured
by predator, $q>0$ the food-independent death rate of predator and
$K>0$ the carrying capacity, $a>0$ the so-called half-saturation
constant, $b>-2\sqrt{a}$ such that the denominator of above system
does not vanish for non-negative $N$.

On the other hand, the real world we lived in is a spatial world,
and spatial patterns are ubiquitous in nature, which modify the
temporal dynamics and stability properties of population density on
range of spatial scales, whose effects must be incorporated in
temporal ecological models that do not represent space
explicitly~\cite{Claudia}. And the spatial component of ecological
interactions has been identified as an important factor in how
ecological communities are shaped. Empirical evidence suggests that
the spatial scale and structure of the environment can influence
population interactions and the composition of communities
~\cite{Cantrell2003}.

Reaction-diffusion model is a typical spatial extended model. It
involves not only time but also space and consists of several
species which react with each other and diffuse within the spatial
domain. It also involves a pair of partial differential equations,
and represents the time course of reacting and diffusing process. In
the spatial extended predator-prey model, the interaction between
the predator and the prey is the reaction item, and the diffusion
item comes into being for the predator's ``pursuit" and the prey's
``evasion". Diffusion is a spatial process, and the whole model
describes the evolution of the predator and the prey going with
time.

Decades after Turing~\cite{Turing1952} demonstrated that spatial
patterns could arise from the interaction of reactions or growth
processes and diffusion, reaction-diffusion models have been studied
in ecology to describe the population dynamics of predator-prey
model for a long time since Segel and Jackson applied Turing's
idea~\cite{Segel1972}. Since then, a new field of ecology, pattern
formation, came into being. The problem of pattern and scale is the
central problem in ecology, unifying population biology and
ecosystem science, and marrying basic and applied
ecology~\cite{Levin1992}. And the study of spatial patterns in the
distribution of organisms is a central issue in ecology, geology,
chemistry, physics and so
on~\cite{Hawick2006,Martin2006,Cantrell2003,RevModPhys.65.851,Garvie,Gierer1972,Griffith,
J.Garcia-Ojalvo,Estep,Klausmeier06111999,Teemu2004,Karen,Maini2002,Maini2006,Maionchi,Malchow2004,medvinsky:311,Murray2003,Koichiro,
Wang2007,Zhou,LiZzh2005,yang:7259,PhysRevLett.88.208303,YangLF2004,Katsuragi,Riaz2005,Riaz2006}.
Theoretical work has shown that spatial and temporal pattern
formation can play a very important role in ecological and
evolutionary systems. Patterns can affect, for example, stability of
ecosystems, the coexistence of species, invasion of mutants and
chaos. Moreover, the patterns themselves may interact, leading to
selection on the level of patterns, interlocking ecoevolutionary
time scales, evolutionary stagnation and diversity.

Based on the above discussions, the spatial extended Holling-type IV
predator-prey model with reaction diffusion takes the form:
\begin{eqnarray}\label{eq:5}
 \begin{array}{l}
 \frac{\partial N}{\partial t}=rN \left( 1-{\frac {N}{K}} \right)
  -{\frac {mNP}{1+bN+a{N}^{2}}}+d_1\nabla^2N,\qquad
 \frac{\partial P}{\partial t}=P \left( -q+{\frac {cmN}{1+bN+a{N}
 ^{2}}} \right)+d_2\nabla^2P,
 \end{array}
\end{eqnarray}
where $d_1$ and $d_2$ are the diffusion coefficients respectively,
$\nabla^2=\frac{\partial}{\partial x^2}+\frac{\partial}{\partial
y^2}$ the usual Laplacian operator in two-dimensional space, and
other parameters are the same definition as those in
model~\eqref{eq07}.

Easy to know that, when a spatial extended predator-prey model is
considered, the evolution of the model is decided by two sorts of
sources (internal source and external source) which act together.
The internal source is the dynamics of the individuals of the model,
and the external source is the variability of environment. Some of
the variability is periodic, such as temperature, water, food supply
of the prey and mating habits. It is necessary and important to
consider models with periodic ecological parameters or perturbations
which might be quite naturally exposed~\cite{Cushing}. These
periodic factors are regarded as the external periodic forcing in
the predator-prey systems. The external forcing can affect the
population of the predator and prey, respectively, which would go
extinct in a deterministic environment. And some of the variability
is irregular, such as the the seasonal changes of the weather, food
supply of the prey, mating habits, and the affects of this
variability are called ``noise''. Ecological population dynamics are
inevitably ``noisy"~\cite{Kendall}. In the predator-prey systems,
the random fluctuations are also undeniably arising from either
environmental variability or internal species. To quantify the
relationship between fluctuations and species' concentration with
spatial degrees of freedom, the consideration of these fluctuations
supposes to deal with noisy quantities whose variance might at time
be a sizable fraction of their mean levels. For example, the birth
and death processes of individuals are intrinsically stochastic
fluctuations which becomes especially pronounced when the number of
individuals is small~\cite{Malchow2004}. Moreover, there are many
other stochastically factors causing predator-prey population to
change, such as effects of spatial structure of the habitat on the
predator-prey ecosystem. The interaction between the predator and
prey, which are far from being uniformly distributed, also introduce
randomness. And these processes can be regarded as parameter
fluctuates irregularly with spaces and time.

The induced effects of the external forcing and noise in population
dynamics, such as pattern formation, stochastic resonance, delayed
extinction, enhanced stability, quasi periodic oscillations and so
on, have been investigated with an increasing interest in the past
decades~\cite{YangLF2004,
J.Garcia-Ojalvo,Jose1998,Spagnolo,Zhou,Sifenni,Liuqx,Mankin2006,Malchow2004,Riaz2005}.
And noise cannot systematically be neglected in models of population
dynamics~\cite{Jose1998}. Zhou and Kurths~\cite{Zhou} concluded
these periodic variabilities as external forcing, and investigated
the interplay among noise, excitability, mixing and external forcing
in excitable media advected by a chaotic flow, in a two-dimensional
FitzHugh-Nagumo model described by a set of
reaction-advection-diffusion equations. And Si et al~\cite{Sifenni}
studied the propagation of traveling waves in sub-excitable systems
driven, and Liu et al~\cite{Liuqx} considered a spatial extended
phytoplankton-zooplankton system with additive noise and periodic
forcing. Following these models they considered, the Holling-type IV
predator-prey model with external periodic forcing and colored noise
is as follows:
\begin{eqnarray}\label{eq:6}
 \begin{array}{l}
 \frac{\partial N}{\partial t}=rN \left( 1-{\frac {N}{K}} \right)
  -{\frac {mNP}{1+bN+a{N}^{2}}}+A\sin{(\omega t)}+d_1\nabla^2N,\\[4pt]
 \frac{\partial P}{\partial t}=P \left( -q+{\frac {cmN}{1+bN+a{N}
 ^{2}}} \right)+\eta (\textbf{\emph{r}},t)+d_2\nabla^2P,
 \end{array}
\end{eqnarray}
where $A\sin(\omega t)$ denotes the periodic forcing with amplitude
$A$ and angular frequency $\omega$. The colored noise term
$\eta(\mathbf{r},t)\,\,\,\,(\,\mathbf{r}=(x, y)\,)$ is introduced
additively in space and time, referring to the fluctuations in the
predator increase rate, which partially results from the
environmental factors such as epidemics, weather and nature
disasters, and it is the Ornstein-Uhlenbech process that obeys the
following linear stochastic partial differential equation
\begin{equation}\label{eq3}
\frac{\partial\eta(\mathbf{r},t)}{\partial t}=-\frac{1}{\tau
}\eta(\mathbf{r},t)+\frac{1}{\tau}\xi(\mathbf{r},t),
\end{equation}
where $\xi(\mathbf{r},t)$ is a Gaussian white noise or the so called
Markovian random telegraph process in both space and time with zero
mean and correlation:
$$
\langle\xi(\mathbf{r},t)\rangle=0, \quad \langle\xi(\mathbf{r},t)
\xi(\mathbf{r}',t')\rangle=2
\varepsilon\delta(\mathbf{r}-\mathbf{r}') \delta(t-t'),
$$
where $\langle\cdot\rangle$ denotes mean value with respect to the
noise $\xi(\mathbf{r},t)$, and $\delta$ the Dirac delta-function,
$\delta(\mathbf{r}-\mathbf{r}')$ the spatial correlation function of
the Gaussian white noise $\xi(\mathbf{r},t)$ .

Integrating equation~\eqref{eq3} with respect to time $t$, we get
$$
\eta (\mathbf{r},t)=\eta (\mathbf{r},0)
e^{-\frac{t}{\tau}}+\frac{1}{\tau} e ^{-\frac{t}{\tau}}\int_0^t e^{
\frac {s}{\tau}}\xi(\mathbf{r},s) d s.
$$
The mean value of the colored noise is
$$
\langle \eta (\mathbf{r},t)\rangle=\langle \eta
(\mathbf{r},0)\rangle e^{
-\frac{t}{\tau}}+\frac{1}{\tau}e^{-\frac{t}{\tau}}\int_0^t e^{\frac
{s}{\tau}}\langle\xi(\mathbf{r},s)\rangle d s=\langle \eta
(\mathbf{r},0)\rangle e^{ -\frac{t}{\tau}},
$$
and the correlation function of the colored noise is given by
\begin{eqnarray}\label{eq8}
\langle \eta (\mathbf{r},t)\eta(\mathbf{r}',t')\rangle&=&\langle
\eta (\mathbf{r},0)\rangle \langle \eta (\mathbf{r}',0)\rangle
e^{-\frac{t+t'}{\tau}}+
\frac{1}{\tau^2}e^{-\frac{t+t'}{\tau}}\int_0^t\int_0^{t'} e^{\frac
{s+s'}{\tau}}\langle\xi(\mathbf{r},s)\xi(\mathbf{r}',s')\rangle
d s d s'  \\
&=&\langle \eta (\mathbf{r},0)\rangle \langle \eta
(\mathbf{r}',0)\rangle e^{
-\frac{t+t'}{\tau}}+\frac{\varepsilon}{\tau^2}e^{
-\frac{t+t'}{\tau}}\delta(\mathbf{r}-\mathbf{r}')\int_0^t\int_0^{t'}
e^{\frac
{s+s'}{\tau}}\delta(t-t') d s d s' \\
&=&\langle \eta(\mathbf{r},0)\rangle \langle \eta
(\mathbf{r}',0)\rangle e^{
-\frac{t+t'}{\tau}}+\frac{\varepsilon}{\tau}\delta(\mathbf{r}-\mathbf{r}')(e^{
-\frac{t+t'}{\tau}}-2e^{ -\frac{t}{\tau}}+e^{-\frac{t-t'}{\tau}}).
\end{eqnarray}
Let $t\rightarrow +\infty$, then
$$
\langle \eta (\mathbf{r},t)\eta(\mathbf{r}',t')\rangle \rightarrow
\frac{\varepsilon}{\tau}e^{-\frac{t-t'}{\tau}}\delta(\mathbf{r}-\mathbf{r}').
$$

The colored noise $\eta (\mathbf{r},t)$ generated in this way
represents a simple spatiotemporal structured noise that can be used
in real mimic situations, which is temporally correlated and white
in space, and satisfies
$$
\langle
\eta(\mathbf{r},t)\eta(\mathbf{r}',t')\rangle=\frac{\varepsilon}{\tau}
e^{-\frac{|t-t'|}{\tau}} \delta(\mathbf{r}-\mathbf{r}'),
$$
where the temporal memory of the stochastic process controlled by
$\tau$ and $\varepsilon$ is the intensity of noise. In this paper,
we set $\tau=1$.

Based on these discussion above, in this paper, we mainly focus on
the spatiotemporal dynamics of model~\eqref{eq:5} and~\eqref{eq:6}.
And the organization is as follows: In section 2, we employ the
method of stability analysis to derive the symbolic conditions for
Hopf and Turing bifurcation in the spatial domain. In section 3, we
give the complex dynamics of model~\eqref{eq:5} and~\eqref{eq:6},
involving pattern formation, phase portraits, time series plots and
resonant response and so on, via numerical simulation. Then, in the
last section, we give some discussions and remarks.

\section{Hopf and Turing bifurcation}

The non-spatial model~\eqref{eq07} has at least two equilibria
(steady states) which correspond to spatially homogeneous equilibria
of the model~\eqref{eq:5} and~\eqref{eq:6}, in the positive
quadrant: $(0,0)$ (total extinct) is a saddle; $(K,0)$ (extinct of
the predator, or prey-only) is a attracting node if $q>{\frac
{cmK}{1+Kb+a{K}^{2}}}$, a saddle if $q<{\frac {cmK}{1+Kb+a{K}^{2}}}$
or a saddle-node if $q={\frac {cmK}{1+Kb+a{K}^{2}}}$. When
$$\begin{array}{l}(a,b,c,m,q,r,K)\in E_1,\,\,\text{here,}\,\,E_1=\{(a,b,c,m,q,r,K)|\, mc>qb, {q}^{2}a<(
mc-qb)^{2},\\ \qquad \sqrt{(mc-qb)^{2}-4\,{q}^{2}a}>{\frac
{-{m}^{2}{c}^{2}+2\,qbmc+qaKmc-{q}^{2}aKb+2\,{q}^{2}a-
{q}^{2}{b}^{2}}{-mc+qb+qaK}}>0,{\frac {b}{a}}-\,{\frac
{mc}{qa}}+\,K<0\},\end{array}$$ there exists unique stationary
coexistent state $({N_1}^*,{P_1}^*)$, where
$$
\begin{array}{l}
{N_1}^*={\frac {1}{2}}\,{\frac {-qb+mc-A}{qa}},\qquad {P_1}^*={\frac
{cr \left( (-mc\,+bq+qaK\,){N_1}^*+q \right) }{a {q}^{2}K}}.
\end{array}$$
On the other hand, when
$$\begin{array}{l}(a,b,c,m,q,r,K)\in E_2,\,\,\text{here,}\,\,E_2=\{(a,b,c,m,q,r,K)|\, mc>qb, {q}^{2}a<(
mc-qb)^{2},\\ \qquad \sqrt{(mc-qb)^{2}-4\,{q}^{2}a}>-{\frac
{-{m}^{2}{c}^{2}+2\,qbmc+qaKmc-{q}^{2}aKb+2\,{q}^{2}a-
{q}^{2}{b}^{2}}{-mc+qb+qaK}}>0,{\frac {b}{a}}-\,{\frac
{mc}{qa}}+\,K>0\},\end{array}$$ there exists another unique
stationary coexistent state $({N_2}^*,{P_2}^*)$ implying:
$$
\begin{array}{l}
{N_2}^*={\frac {1}{2}}\,{\frac {-qb+mc+A}{qa}},\qquad {P_2}^*={\frac
{cr \left( (-mc\,+bq+qaK\,){N_2}^*+q \right) }{a {q}^{2}K}}.
\end{array}$$

It is worth mentioning that equilibria $(N_1^*, P_1^*)$ and $(N_2^*,
P_2^*)$ cannot coexist. In this paper, we mainly focus on the
dynamics of $(N_1^*, P_1^*)$ and rewrite it as $(N^*, P^*)$. The
dynamic behavior of $(N_2^*, P_2^*)$ is similar to those of $(N_1^*,
P_1^*)$.

To perform a linear stability analysis, we linearize
model~\eqref{eq07} around the stationary state $(N^*,P^*)$ for small
space- and time-dependent fluctuations and expand them in Fourier
space
$$
N(\mathbf{r},t)\sim N^*e^{\lambda
t}e^{i\vec{k}\cdot\mathbf{r}},\quad P(\mathbf{r},t)\sim
P^*e^{\lambda t}e^{i\vec{k}\cdot\mathbf{r}},\quad
\mathbf{r}=(x,y),\quad \vec{k}=(k_x, k_y).
$$
where $\lambda$ is the eigenvalue of the Jacobian matrix of
model~\eqref{eq07}.

Hopf bifurcation is an instability induced by the transformation of
the stability of a focus. Mathematically speaking, Hopf bifurcation
occurs when $ \text{Im}(\lambda)\neq 0, \quad
\text{Re}(\lambda)=0\,\,\,\text{at}\,\,\,k=0$, where
$\text{Im}(\lambda)$ is the imaginary part, $\text{Re}(\lambda)$ the
real part and $k$ the wave number. So we get the Hopf bifurcation
surface
\begin{equation}\label{eq:7}
H=\{(a,b,c,m,q,r,K)|\,\,\text{det}(J_0)> 0,\text{trace}(J_0)=0\},
\end{equation}
where
$$\begin{array}{l}
\text{det}(J_0)=-\left( r-2\,{\frac {rN^*}{K}} \right) q+ \frac{
mqP^*+  cm\left( r-2\,{ \frac {rN^*}{K}} \right)N^*
 }{ \left(
1+bN^*+a{N^*}^{2} \right)}  -{\frac {mqN^*P^* \left( b+2\,aN^*
\right) }{ \left(
1+bN^*+a{N^*}^{2} \right) ^ {2}}},\\[8pt]
\text{trace}(J_0)=r-2\,{\frac {rN^*}{K}}-q +{\frac {m \left(
-P^*+cN^*+a{N^*}^{2}P^*+bc{N^*}^{2}+c{N^*} ^{3}a \right) }{ \left(
1+bN^*+a{N^*}^{2} \right) ^{2}}},
 \end{array}
$$
the frequency of periodic oscillations in time $\omega_H$ satisfies
$\omega_H=\text{Im}(\lambda)=\sqrt{\text{det}(J_0)}$, and the
corresponding wavelength $\lambda_H$ satisfies $
\lambda_H=\frac{2\pi}{\omega_H}=\frac{2\pi}{\sqrt{\text{det}(J_0)}}\,.$
Especially, we take $K$ as the bifurcation parameter, and can get
the critical value of Hopf bifurcation from Eq.~\eqref{eq:7}:
\begin{equation}\label{eq:8}
\begin{array}{l}
K_H=(-(a{q}^{2}(5\,mc-3\,qb)-(3\,mc-qb
)(mc-qb)^{2})\sqrt {(mc-qb)^{2}-4\,{q}^{2}a}\\[8pt]-4\,{q}^{4}{a}^{2} +{q}^{2}
(mc-qb)(11\,mc-5\,qb)a-(3\,mc-qb)(mc-qb)^{3} )
/(-aq(((2\,mc-qb)\\[8pt](mc-qb)-2\,{q}^{2}a)\sqrt
{(mc-qb)^{2}-4\,{q}^{2}a}-2a{q}^{2}(3\,mc-2\,qb)
+(2\,mc-qb)(mc-qb)^{2})).
\end{array}
\end{equation}

Turing instability is induced only by ``pursuit and evasion"  if the
predator can catch the prey by pursuit. We call the critical state
of Turing instability as Turing bifurcation. Turing bifurcation
occurs when $\text{Im}(\lambda)=0, \,\,
\text{Re}(\lambda)=0\,\,\text{at}\,\,k=k_T\neq0,$ and the wavenumber
$k_T$ satisfies $k_T^2=\sqrt{\frac{\text{det}(J_0)}{d_1d_2}}$. In
addition, at the Turing threshold, the spatial symmetry of the
system is broken and the patterns are stationary in time and
oscillatory in space with the wavelength
$\lambda_T=\frac{2\pi}{k_T}$. And the Turing bifurcation surface
implies:
\begin{equation}\label{eq:9}
T=\{(a,b,c,m,q,r,d_1,d_2,K)|\quad\text{det}(J_k)=
0,\text{trace}(J_k)=0\},
\end{equation}
where
$$
\begin{array}{l}
\text{det}(J_k)=- \left( r-2\,{\frac {rN^*}{K}}-d_{{1}}{k}^{2}
\right) \left( q+d_{{2}} {k}^{2}\right)\frac{\left(  \left( q
+d_{{2}}{k}^{2} \right)mP^* + \left( r -2\,{\frac
{rN^*}{K}}-d_{{1}}{k}^{2} \right) cmN^* \right)}
{1+bN^*+a{N^*}^{2}}\\[8pt]\qquad\qquad\,\,\,-{ \frac {m \left( b+2\,aN^* \right) \left(
q+d_{{2}}{k}^{2} \right)N^*P^* }{( 1+bN^*+a{N^*}^{2})^{2}}},\\[8pt]
\text{trace}(J_k)=r-2\,{\frac {rN^*}{K}}-q- \left( d_{{1}}+d_{{2}}
\right){k}^{2} +{\frac {m \left(
-P^*+cN^*+a{N^*}^{2}P^*+bc{N^*}^{2}+c{N^*} ^{3}a \right) }{ \left(
1+bN^*+a{N^*}^{2} \right) ^{2}}},
\end{array}
$$
the critical value of Turing bifurcation can be obtained from
Eq.~\eqref{eq:9} as follows:
\begin{equation}
K_{{T}}={\frac {F_{{1}}}{F_{{2}}}},
\end{equation}
where
\begin{widetext}
$${
\begin{array}{l}
F_{{1}}=r((4\,{q}^{2}a(2\,mc-qb)-(3\,mc-qb)(mc-qb)^{2})\sqrt
{(mc-qb)^{2}-4\,{q}^{2}a}+((mc-qb)^{2}-4\,{q}^{2}a)\\[4pt]
\qquad\qquad\cdot((3\,mc-qb )(mc-qb) -2\,{q}^{2}a ))
((3\,mc-qb)(mc-qb)-4\,{q}^{2}a\\[4pt]
\qquad\qquad-(3\,mc+qb)
\sqrt {(mc-qb)^{2}-4\,{q}^{2}a}\,\,)d_{{2}},\\[8pt]
\end{array}}$$
$${
\begin{array}{l}
F_{{2}}=qa(2\,mc((mc-qb)\sqrt
{(mc-qb)^{2}-4\,{q}^{2}a}+4\,{q}^{2}a-(mc-qb )^{2}) B+
((3\,mc-qb)d_{{2}}\\[4pt]
\qquad\qquad\cdot(2\,{m}^{2}{c}^{2}-3\,qbmc+{q}^{2}{b}^{2}-4\,{q}^{2}a)r
+2\,qd_{{1}}mc(mc-qb)^{2})({(mc-qb)^{2}-4\,{q}^{2}a})\\[4pt]
\qquad\qquad +(-2\,d _{{2}}(( 3\,mc-qb )(
2\,mc-qb)(mc-qb)^{2}-2\,{q}^{2}a(-4\,{q}^{2}a+3\,{q}^{2}{
b}^{2}-12\,qbmc\\[4pt]
\qquad\qquad+11\,{m}^{2}{c}^{2}))r-4\,qcm(mc-qb)d_{{1}}((
mc-qb)^{2}-4\,{q}^{2}a))\sqrt{(mc-qb)^{2}-4\,{q}^{2}a}\\[4pt]
\qquad\qquad+( {q}^{2}{b}^{2}-2\,qbmc+{m}^{2}{c}^{2}-4\,{
q}^{2}a)((2\,mc-qb)
d_{{2}}(3\,{m}^{2}{c}^{2}-4\,qbmc+{q}^{2}{b}^{2}-4\,{q}^{2}a)
r\\[4pt]
\qquad\qquad+2\,mcqd_{{1}}((mc-qb)^{2}-4\,{q}^{2}a))
),\\[8pt]
B=(-2\,d_{{1}}q(((d_{{1}}b{q}^{2}-qd_{{1}}m
c-qrbd_{{2}}+2\,rmcd_{{2}})(4\,{q}^{2}a-(mc-qb
)^{2})-rmcd_{{2}}(mc-qb)^{2})\\[4pt]
\qquad\qquad\cdot\sqrt{(mc-qb)^{2}-4\,{q}^{2}a}+(
8\,{q}^{4}d_{{2}}r-8\,{q}^{5}d_{{1}}){a}^{2}+4\,(mc-qb)
a{q}^{2}rmcd_{{2}}+(3\,rmcd_{
{2}}\\[4pt]
\qquad\qquad -qd_{{1}}mc+d_{{1}}b{q}^{2}-qrbd_{{2}})((
mc-qb)^{3}-6\,(mc-qb)a{q}^{2})) )^{\frac{1}{2}}.
\end{array}}$$
\end{widetext}

\begin{figure*}[htp]
\includegraphics[width=16cm,height=9cm]{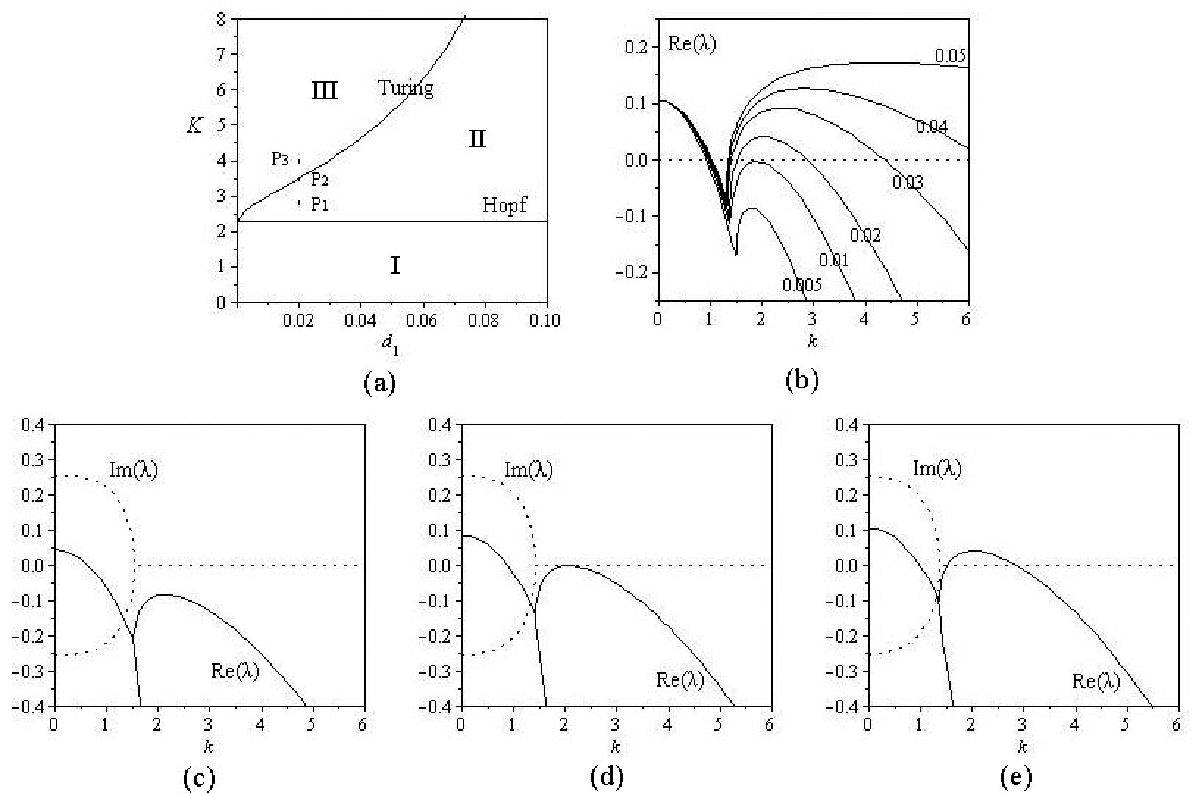}
\caption{\label{fig1} (a) $K-d_1$ Bifurcation diagram for model
\eqref{eq:5} with $r=1$, $a=0.125$, $b=1$, $c=0.7$, $m=0.625$,
$q=0.18$ and $d_2=0.2$. Hopf and Turing bifurcation lines separate
the parameter space into three domains. $P_1 (0.02, 2.8)$, $P_2
(0.02, 3.499)$, $P_3 (0.02, 4.0)$ are the different selections
corresponding to (c), (d), (e), respectively. (b) Dispersion
relations showing unstable Hopf mode, transition of Turing mode from
stable to unstable for the system model \eqref{eq:5} as $d_1$
decreased. The other parameters in (c)--(e) are: $d_1=0.02$, the
bifurcation parameter $K$ equals: (c) $2.8>K_H=2.279$; (d)
$3.499=K_T$; (e) $4.0>K_T>K_H$. The real parts $\text{Re}(\lambda)$
and the imaginary parts $\text{Im}(\lambda)$ are shown by solid
curves and dashed curves, respectively. }
\end{figure*}

Linear stability analysis yields the bifurcation diagram with $r=1$,
$a=0.125$, $b=1$, $c=0.7$, $m=0.625$, $q=0.18$ and $d_2=0.2$ shown
in Figure~\ref{fig1}(A). In this case, parameters
$(a,b,c,m,q,r,K)\in E_1$, and $(N^*,P^*)$ is the unique stationary
coexistent state. From Figure~\ref{fig1}(a), one can see that the
Hopf bifurcation line and the Turing bifurcation curve separate the
parametric space into three distinct domains. In domain I, located
below all two bifurcation lines, the steady state is the only stable
solution of the model. Domain II is the region of pure Hopf
instability. When the parameters correspond to domain III, located
above all two bifurcation lines, both Hopf and Turing instability
occur. Figure~\ref{fig1}(b) displays the dispersion relations
showing unstable Hopf mode, transition of Turing mode from stable to
unstable for model~\eqref{eq:5}, e.g., as $d_1$ decreased.
Figure~\ref{fig1}(c) illustrates the relation between the real and
the imaginary parts of the eigenvalue $\lambda$ with
$K=2.8>K_H=2.279$, located in domain II, one can see that when
$k=0$, $\text{Re}(\lambda(k))>0$ and $\text{Im}(\lambda(k))\neq 0$.
Figure~\ref{fig1}(d) displays the case of the critical value of
Turing bifurcation $K=K_T=3.499$, in this case,
$\text{Re}(\lambda(k))=0$ and $\text{Im}(\lambda(k))=0$ at
$k=k_T=2.080$. When $K=4.0$, located in domain III,
figure~\ref{fig1}(e) indicates that at $k=0$,
$\text{Re}(\lambda(k))>0$, $\text{Im}(\lambda(k))\neq 0$.

\section{Spatiotemporal dynamics of the models}

In this section, we perform extensive numerical simulations of the
spatially extended model~\eqref{eq:5} and~\eqref{eq:6} in
two-dimensional spaces, and the qualitative results are shown here.
All our numerical simulations employ the Zero-flux Neumann boundary
conditions with a system size of $200\times200$ space units. The
parameters are $r=1$, $a=0.125$, $b=1$, $c=0.7$, $m=0.625$,
$q=0.18$, $d_1=0.02$, $d_2=0.2$ and $K=2.8$ or $K=4.0$, which
satisfy $(a,b,c,m,q,r,K)\in E$. Model~\eqref{eq:5} and~\eqref{eq:6}
are integrated initially in two-dimensional space from the
homogeneous steady state, i.e., we start with the unstable uniform
solution $(N^*, P^*)$ with small random perturbation superimposed,
in each, the initial condition is always a small amplitude random
perturbation $(\pm 5\times 10^{-4})$, using a finite difference
approximation for model~\eqref{eq:5} or Fourier transform method for
model~\eqref{eq:6} for the spatial derivatives and an explicit Euler
method for the time integration with a time stepsize of $\Delta
t=1/24$ and space stepsize (lattice constant) $\Delta x=\Delta y=1$.
We have taken some snapshots with white (black) corresponding to the
high (low) value of prey $N$.

In the numerical simulations, different types of dynamics are
observed and we have found that the distributions of the predator
and prey are always of the same type. Consequently, we can restrict
our analysis of pattern formation to one distribution. In this
section, we show the distribution of prey $N$, for instance.

\subsection{Pattern formation of model (4)}

\begin{figure*}[htp]
\includegraphics[width=13cm]{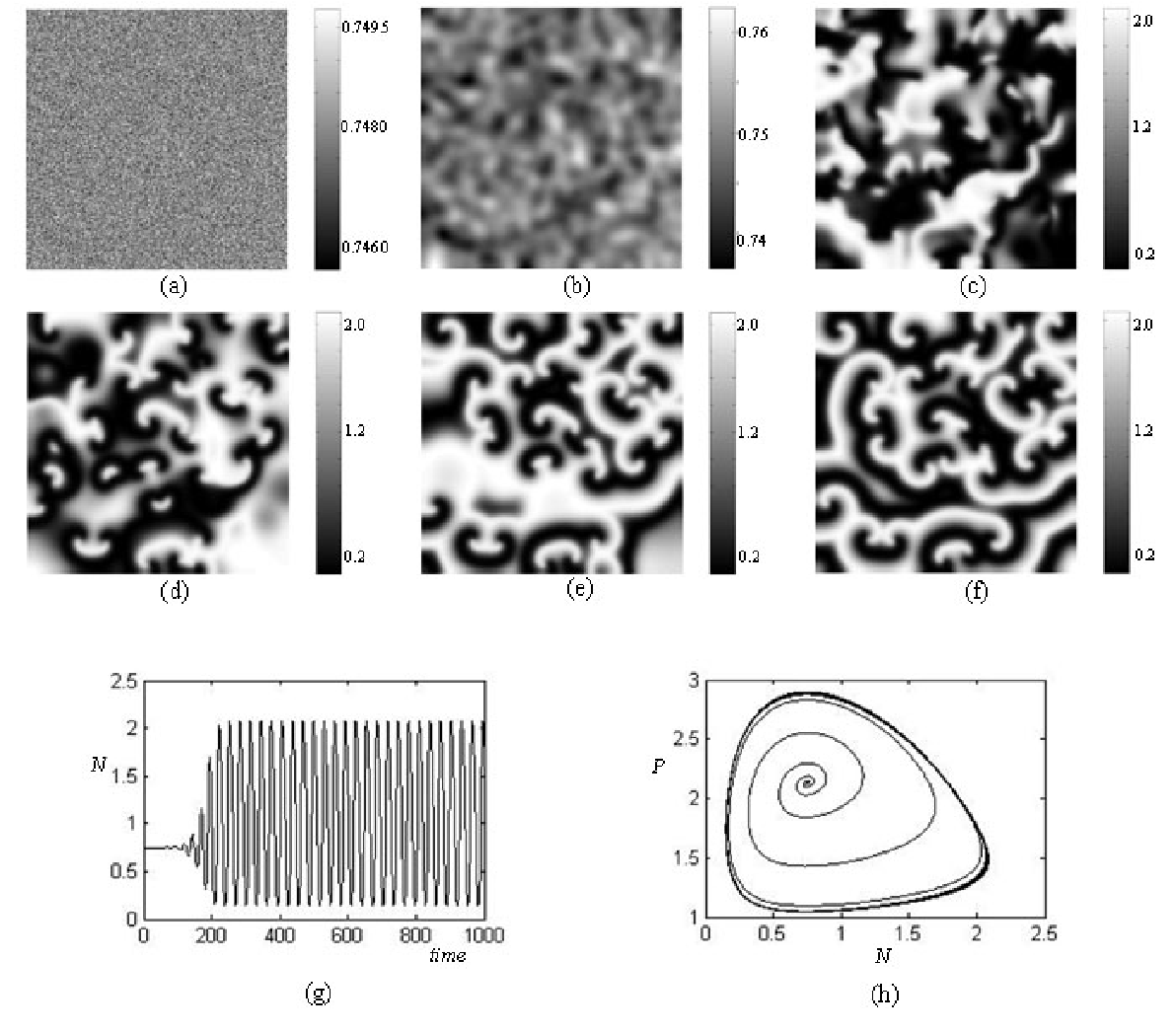}
\caption{\label{fig2} Grey-scaled snapshots of spatiotemporal
pattern of the prey $N$ of model~\eqref{eq:5} with $K=2.8$. (a)
$t=0$, (b) $t=100$, (c) $t=300$, (d) $t=500$, (e) $t=1000$, (f)
$t=2000$. The lower panels show the corresponding (g) time series
plots and (h) phase portraits.}
\end{figure*}

Figure~\ref{fig2} shows the evolution of the spatial patterns of
prey $N$ at $t=0$, 100,  300, 500, 1000, 2000, with random small
perturbation of the equilibrium $(N^*, P^*)=(0.748, 2.132)$ of
model~\eqref{eq:5} with $K=2.8$, located in domain II, more than the
Hopf bifurcation threshold $K_H=2.279$ and less than the Turing
bifurcation threshold $K_T=3.499$. In this case, pure Hopf
instability occurs. One can see that for model~\eqref{eq:5}, the
random initial distribution (c.f., figure~\ref{fig2}(a)) leads to
the formation of macroscopic spiral patterns (c.f.,
figures~\ref{fig2}(d) to (f)). In other words, in this situation,
spatially uniform steady-state predator-prey coexistence is no
longer. Small random fluctuations will be strongly amplified by
diffusion, leading to nonuniform population distributions. From the
analysis in section 2, we find with these parameters in domain II,
the spiral pattern arises from Hopf instability. The lower panel in
figure~\ref{fig2} shows the corresponding (g) time series and (h)
phase portraits. Figure~\ref{fig2}(g) illustrates the evolution
process of prey $N$, periodic oscillating in time finally, (h)
exhibits that a limit cycle arises, which is caused by the Hopf
bifurcation.

\begin{figure*}[htp]
\includegraphics[width=13cm]{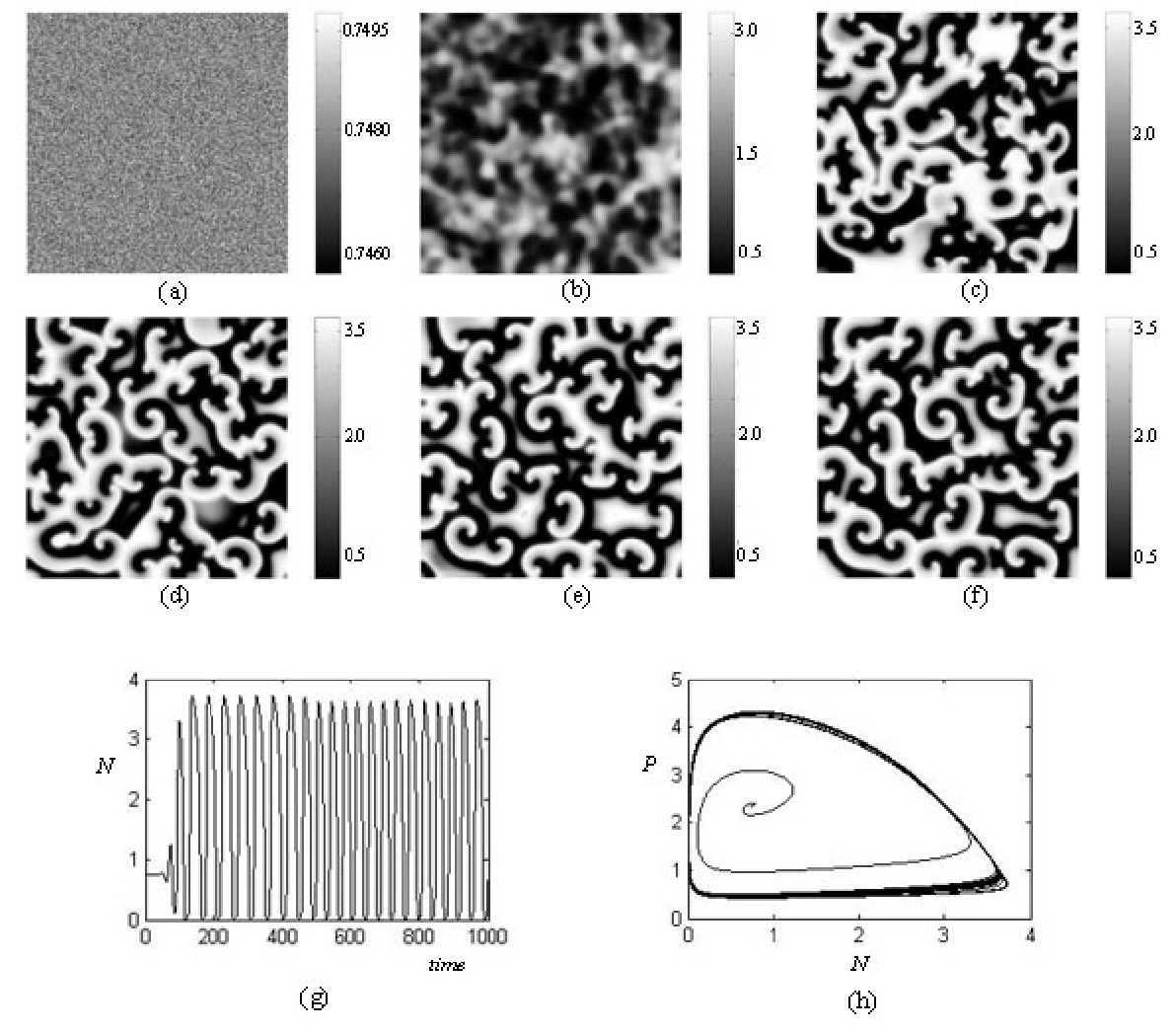}
\caption{\label{fig3} Grey-scaled snapshots of spatiotemporal
pattern of the prey $N$ of model~\eqref{eq:5} with $K=4.0$. (a)
$t=0$, (b) $t=100$, (c) $t=300$, (d) $t=500$, (e) $t=1000$, (f)
$t=2000$. The lower panels show the corresponding (g) time series
plots and (h) phase portraits.}
\end{figure*}

When $K=4.0>K_T>K_H$, in this case, parameters in domain III
(figure~\ref{fig1}(A)), both Hopf and Turing instabilities occur.
The nontrivial stationary state is $(N^*, P^*)=(0.748, 2.365)$. As
an example, the formation of a regular macroscopic two-dimensional
spatial pattern is shown in figure~\ref{fig3}. The lower panel in
figure~\ref{fig3} shows the corresponding (g) time series plots and
(h) phase portraits.

Comparing this situation (figure~\ref{fig3}) with the one above
(figure~\ref{fig2}), easy to see that the pattern formations are all
spiral wave. From the analysis in section 2, we know that when
$K=2.8$, the wavelength $\lambda=3.100$ while at $K=4.0$,
$\lambda=3.021$. And the frequency of periodic oscillations in time
is as inverse proportion to wavelength, so we can know that Turing
instability has a positive effect on the frequency and negative
effect on wavelength. This is the reason why the spiral curves are
more dense in figure~\ref{fig3}(f) than in figure~\ref{fig2}(f). On
the other hand, one can see that when $K=4.0$, the time series plots
(c.f., figure~\ref{fig3}(g)) indicate that when Turing instability
occurs, the solution of model~\eqref{eq:5} is strongly oscillatory
in time while with $K=2.8$ (pure Hopf bifurcation emerges) it is
periodic (c.f., figure~\ref{fig2}(g)). In addition, comparing
figure~\ref{fig2}(g) with figure~\ref{fig3}(g), one can see that
Turing instability has a positive effects on the amplitude of prey
$N$. And from figure~\ref{fig3}(h), one can see that a quasi limit
cycle emerges while in figure~\ref{fig2}(h), it is a cycle. Although
there are some different points between figure~\ref{fig2} and
figure~\ref{fig3}, but we can know that Turing instability can't
give birth to different type patterns. In our previous
work~\cite{Wang2007}, we find that Turing instability can change
pattern type. This may be a very point between the Holling-type IV
with Michael-Menton functional response of predator-prey model.

On the other hand, the basic idea of diffusion-driven instability in
a reaction-diffusion system can be understood in terms of an
activator-inhibitor system or predator-prey model~\eqref{eq:5}. The
functioning of this mechanism is based on three
points~\cite{David2002}. First, a random increase of activator
species (prey, $N$) should have a positive effect on the creation
rate of both activator (prey, $N$) and inhibitor (prey, $P$)
species. Second, an increment in inhibitor species should have a
negative effect on formation rate of both species. Finally,
inhibitor species $P$ must diffuse faster than activator species
$N$. Certainly, the reaction-diffusion predator-prey
model~\eqref{eq:5}, with Holling-type IV functional response and
predators diffusing faster than prey (i.e., $d_2>d_1$), provides
this mechanism. And spirals and curves are the most fascinating
clusters to emerge from the predator-prey model. A spiral will form
from a wave front when the prey line (which is leading the front)
overlaps the pursuing line of predator~\cite{Hawick2006}. The prey
on the extreme end of the line stops moving as there is no predator
in the immediate vicinity. However, the prey $N$ and the predator
$P$ in the center of the line continue moving forward. This forms a
small trail of prey at one (or both) ends of the front. These prey
start breeding and the trailing line of prey thickens and attracts
the attention of predator at the end of the fox line that turn
towards this new source of prey. Thus a spiral forms with predator
$P$ on the inside and prey $N$ on the outside. If the original
overlap of prey occurs at both ends of the line a double spiral will
form. Spirals can also form as a prey blob collapses after the
predator eats into it. This is the reason why the pattern formation
of model~\eqref{eq:5} is spiral wave.

\subsection{The effect of noise only of model (5) }

Now, we turn our focus on the effect of noise on the predator $P$ of
stochastic model~\eqref{eq:6}. In this case, $A=0$, i.e., the
periodic forcing is not at presence.

\begin{figure*}[htp]
\includegraphics[width=13cm]{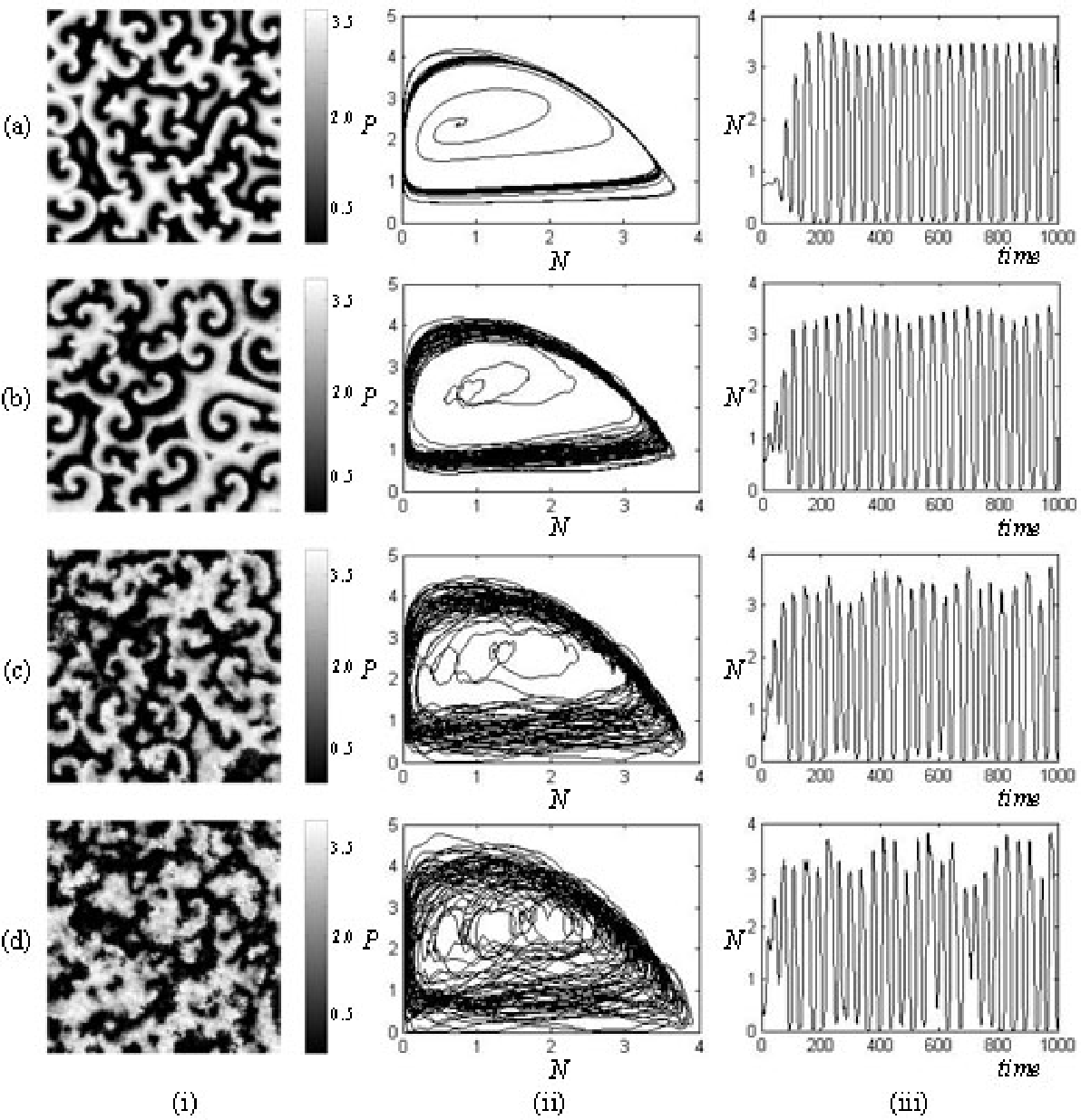}
\caption{\label{fig4} Dynamics of model~\eqref{eq:5}, for the
following noise intensity. (a) $\varepsilon=0.0001$;
(b)$\varepsilon=0.01$; (c)$\varepsilon=0.05$; (d)$\varepsilon=0.1$.
(i) snapshots of pattern formation at time $2000$; (ii) phase
portraits; (iii) time-series plots. $A=0$ and the other parameters
are the same as those in figure~\ref{fig2}.}
\end{figure*}

Figure~\ref{fig4} shows the dynamics of model~\eqref{eq:6} with
noise on the predator. The first row of figure~\ref{fig4}, i.e.,
(a), $\varepsilon=0.0001$; the second row, (b), $\varepsilon=0.01$;
the third row, (c), $\varepsilon=0.05$; and the last row of
figure~\ref{fig4}, (d), $\varepsilon=0.1$. And the first collum of
figure~\ref{fig4}, marked as (i), shows the snapshots of
spatiotemporal pattern of model~\eqref{eq:6} at $t=2000$ with
different intensity of noise, respectively. In this case, one can
see that the pattern formation turns into spatial chaotic from
spiral wave with the increase of noise intensity $\varepsilon$. And
the second collum of figure~\ref{fig4}, marked as (ii), displays the
phase portraits of model~\eqref{eq:6} with different intensity of
noise, respectively. We can see that, as noise intensity
$\varepsilon$ increasing, the symmetry of the limit cycle is broken
and gives rise to chaos. The last collum of figure~\ref{fig4},
(iii), illustrates the time-series plots of prey $N$ with different
intensity of noise, respectively. One can see that noise breaks the
periodic oscillations in time and gives rise to drastically ruleless
oscillations in time.

\subsection{The effect of periodic forcing of model (5)}

In the previous subsection, we have shown the effect of noise on the
predator $P$ of model~\eqref{eq:6}. An interesting question is
whether such noise-sustained oscillations can be entrained by a weak
external forcing, in this case, $\varepsilon=0$, which is
investigated here.

When model~\eqref{eq:6} is noise free, there is a phenomenon of
frequency locking or resonant
response~\cite{Liuqx,Mankin2006,Zhou,Sifenni}. That is, without
noise, the spatially homogeneous oscillation does not respond to the
external periodic forcing when the amplitude $A$ is below a
threshold whose value depends on the external periodic
$T_{in}=\frac{2 \pi}{\omega}$. Above the threshold,
model~\eqref{eq:6} may produce oscillations about period $T_{out}$
with respect to external period $T_{in}$, which is called frequency
locking or resonant response. That is, the model produces one spike
within each of the $M=\frac{T_{out}}{T_{in}}$ periods of the
external force, called $M:1$ resonant response~\cite{Zhou,Sifenni}.
The phenomenon of coherent resonance is of great
importance~\cite{Mankin2006}. Following Si~\cite{Sifenni}, in the
present paper, the output periodic $T_{out}$ is defined as follows:
$T_i$ is the time interval between the $i$th spike and $(i+1)$th
spike. $m$ spikes are taken into account and the average value of
them is $T_{out}=\sum\limits_{i=1}^{m}T_i/(m-1)$.

\begin{figure*}[htp]
\includegraphics[width=11.2cm]{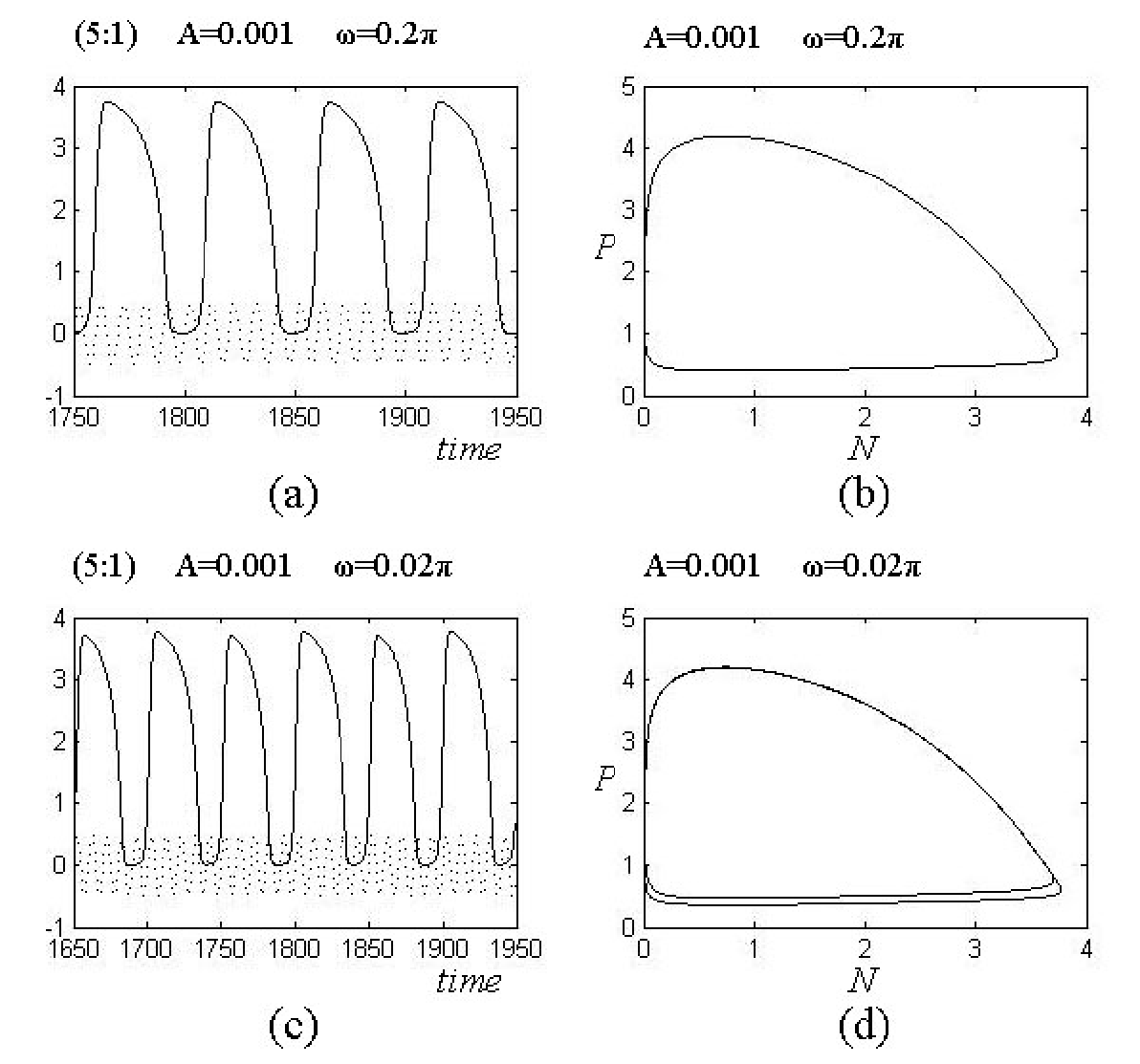}
\caption{\label{fig5} External periodic forcing induced frequency
locking of model\eqref{eq:5}. The solid curve is time series of prey
$U$, the dash curve is the corresponding external periodic forcing.
Other parameters are the same as those in figure~\ref{fig3}. }
\end{figure*}

As an example, with the amplitude $A=0.001$, figure~\ref{fig5} shows
$5:1$ resonant response with $\omega=0.2\pi$ (a) and
$\omega=0.02\pi$ (c), respectively. And figure~\ref{fig5}(b) and (d)
are the phase portraits corresponding to (a) and (c). We can see
that when $\omega=0.2\pi$, there exists a periodic orbit, while
$\omega=0.02\pi$, a periodic-2 orbit of model~\eqref{eq:6} emerges.
Obviously, different $\omega$ can form the same resonant response,
and different phase orbits, i.e., different numerical solution of
model~\eqref{eq:6}, may correspond to the same resonant response.

\subsection{The effect of noise and periodic forcing of model (5)}

\begin{figure*}[htp]
\includegraphics[width=9cm]{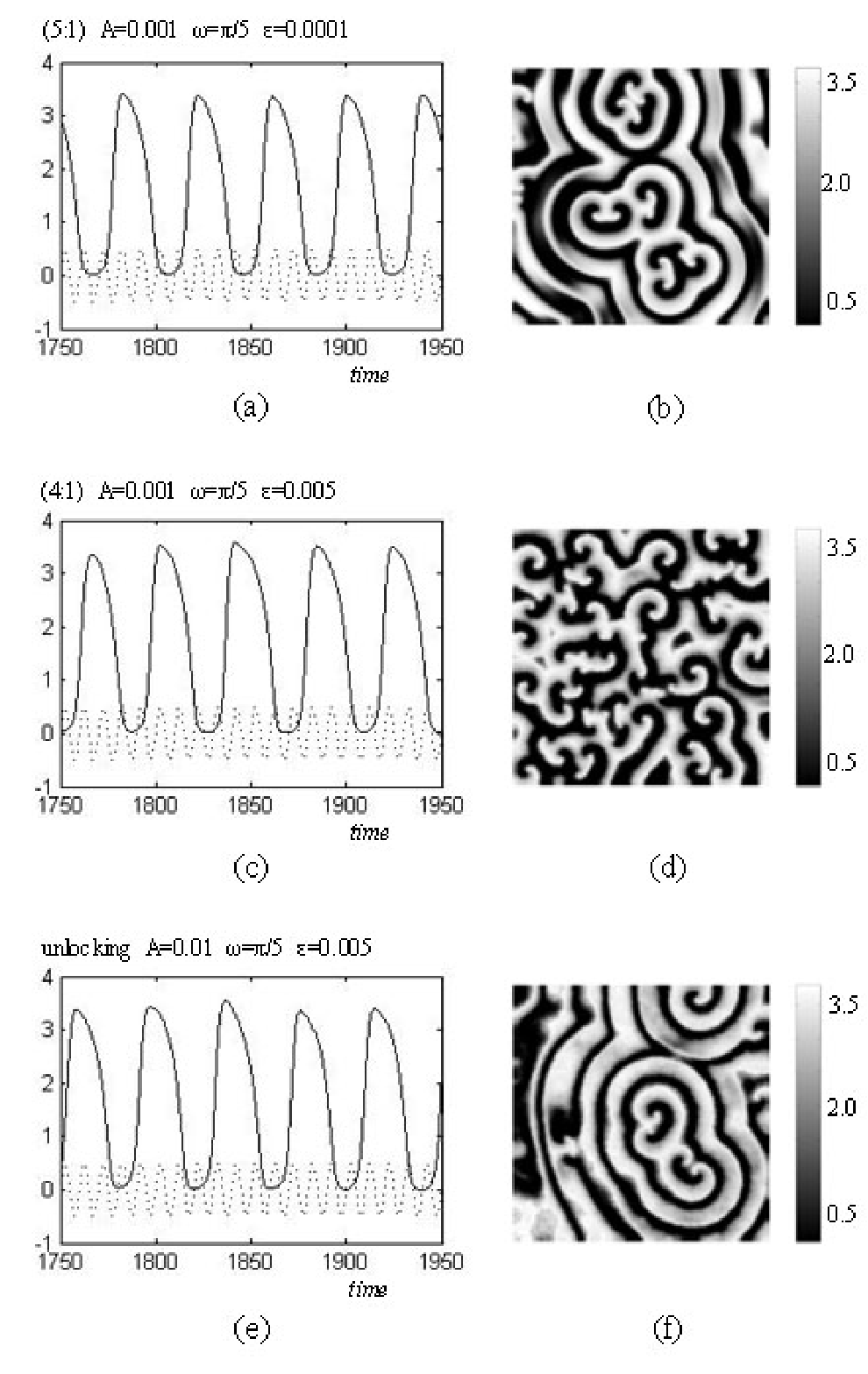}
\caption{\label{fig6} Dynamics of model~\eqref{eq:6} with both noise
and periodic forcing. (b)(d)(f) are snapshots at $t=2000$
corresponding to the left hand side resonant response. The other
parameters are the same as those in figure~\ref{fig3}.}
\end{figure*}

Now, we consider the dynamics about resonant response of
model~\eqref{eq:6} with both noise and periodic forcing. As depicted
in figure~\ref{fig6}, the prey can generate $5:1$ (a), $4:1$ (c)
locked oscillations, depending on the amplitude $A$ and angular
frequency $\omega$. Figures~\ref{fig6} (b) and (d) illustrate the
spiral pattern at $t=2000$ corresponding to (a) and (c),
respectively. For contrast, we change one of the parameters of
figure~\ref{fig6}(c) $A=0.001$ to $A=0.01$ (e), one can see that the
resonant response vanishes, the corresponding spiral pattern (f) is
similar to (b). It indicates that the amplitude $A$ is a control
factor for pattern formation. In addition, comparing
figures~\ref{fig6}(b) with (d), one can see that the pattern
formations are determined by noise intensity $\varepsilon$, too.

\begin{figure*}[htp]
\includegraphics[width=11cm]{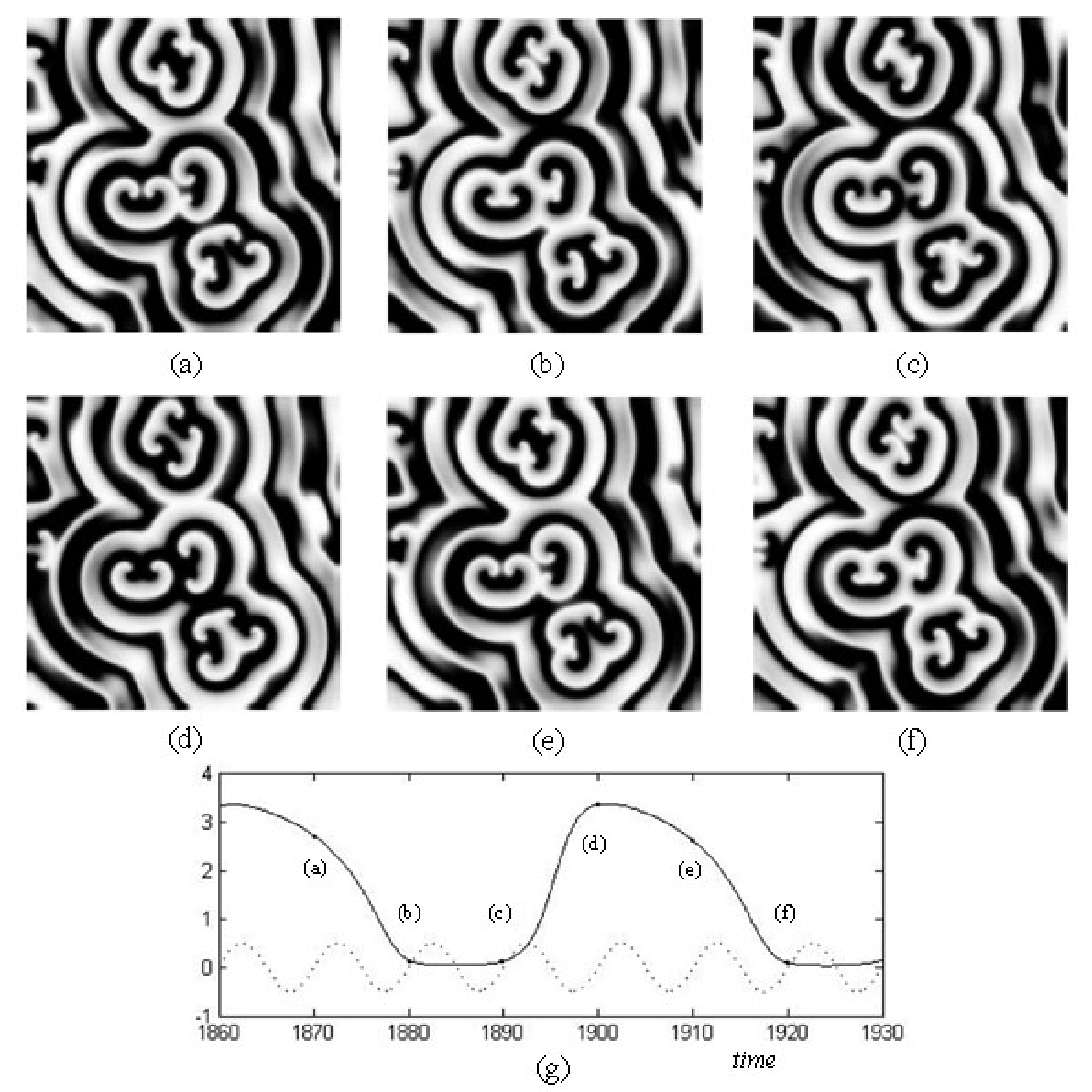}
\caption{\label{fig7} Typical pattern formation of the forced noisy
prey in the $5:1$ locking region at $A=0.001$ and
$\varepsilon=0.0001$ corresponding to figure~\ref{fig6}(a). The
lower panel shows the time series of the prey $N$ (the solid curve)
and the corresponding external periodic forcing (the dash curve)
corresponding to the snapshots of the patterns. The grey scale, from
black to white, is in $[0, 3.5]$ for all the snapshots.}
\end{figure*}

In Figure~\ref{fig7}, we have shown a typical pattern formation
process in the $5:1$ frequency locking regime with $A=0.001$ and
$\omega=0.2\pi$. From $t=1870$ (a) to $t=1920$ (f), the pattern
formation of prey $N$ is spiral wave and some small excitations
already develop. One can see that, during the second periodic of the
forcing, the prey is almost fully synchronized and relaxes slowly
back to the state at moment (f). Obviously, the external periodic
forcing at moment (e) repeats that at moment (a). However, the prey
$N$ does not exactly repeat due to a small fluctuation of the phase
difference.

\section{Conclusions and remarks}

In this paper, we present spatial Holling-type IV predator-prey
model containing some important factors, such as noise (random
fluctuations), the external periodic forcing and diffusion
processes. And the numerical simulations were consistent with the
predictions drawn from the bifurcation analysis, that is, Hopf
bifurcation and Turing bifurcation.

If the parameter $K$, the carrying capacity, located in the domain
II of figure~\ref{fig1}(a), the Hopf instability occurs, the
destruction of the pattern begins from the prey $N$, while it begins
from the predator $P$ if $K$ located in the domain III, both Hopf
and Turing instabilities occur.

Furthermore, we demonstrate that noise and the external periodic
forcing play a key role in the predator-prey model~\eqref{eq:6} with
the numerical simulations. We provoke qualitative transformations of
the response of the model by changing noise intensity, and noise can
enhance the oscillation of the species density, and format large
clusters in the space. The periodic oscillations appear when the
spatial noise and external periodic forcing are turned on. It has
also been realized that model~\eqref{eq:6} is very sensitive to
external periodic forcing through the natural annual variation of
prey growth. In conclusion, we have shown that the cooperation
between noise and external periodic forcing inherent to the
deterministic dynamics of periodically driven models gives rise to
the appearance of resonant response.

Significantly, model~\eqref{eq:6} exhibits oscillations when both
noise and external forcing are present. This means that the predator
population may be partly due to the external forcing and stochastic
factors instead of deterministic factors. Therefore, the model for
spatially extended systems composed by two species could be useful
to explain spatiotemporal behavior of populations whose dynamics is
strongly affected by noise and the environmental physical variables,
and the results of this paper are an important step toward providing
the theoretical biology community with simple practical numerical
methods, for investigating the key dynamics of realistic
predator-prey models.

\begin{acknowledgments}{This work was supported by the National Natural
Science Foundation of China (10471040) and the Youth Science
Foundation of Shanxi Provence (20041004).}
\end{acknowledgments}

\end{document}